# Black Sun: Ocular Invisibility of Relativistic Luminous Astrophysical Bodies

Jeffrey S. Lee[1]
Gerald B. Cleaver[1,2]

[1]Early Universe Cosmology and Strings Group,
Center for Astrophysics, Space Physics, and Engineering Research
[2]Department of Physics
Baylor University
One Bear Place
Waco, TX 76706

Jeff_Lee@Baylor.edu
Gerald_Cleaver@Baylor.edu

**Abstract**

The relativistic Doppler shifting of visible electromagnetic radiation to beyond the human ocular range reduces the incident radiance of the source. Consequently, luminous astrophysical bodies (LABs) can be rendered invisible with sufficient relativistic motion. This paper determines the proper distance as a function of relativistic velocity at which a luminous object attains ocular invisibility.

## 1. Introduction

The relativistic blackbody spectrum [1] suggests the intriguing possibility that a luminous astrophysical body can be rendered optically invisible by Doppler shifting the wavelengths of maximum intensity from the visible frequency range to above or below the frequency thresholds of human vision.

Also, relativistic blackbody radiators will emit spectral radiances which are increased (in the case of approaching) or decreased (in the case of receding), due to temperature inflation and relativistic beaming. By considering the relativistic blackbody spectrum, the proper distances are determined at which the apparent magnitude of a blackbody radiator is greater (i.e. dimmer) than approximately 6.5 (the threshold of vision for the typical unaided human eye).

Additionally, laboratory tests of the sensitivity of the unassisted human eye are described, and this paper asserts that the Judd & Voss CIE 1978 photopic luminous efficiency function would not be applicable to the situation of LABs due to the much greater luminosity than in the laboratory tests.

## 2. The Apparent Magnitude of Blackbody Radiators in the Rest Frame

The relationship between absolute magnitude, apparent magnitude, and distance to an arbitrary stationary blackbody radiation source has been well established and is given by:

$$M = m + 5 - 5\log z \quad (1)$$

where $M$ is the absolute magnitude of any blackbody radiator, $m$ is its apparent magnitude, and $z$ is the distance to the observer in parsecs. Also, in terms of luminosity,

$$M = M_o - 2.5\log\left(\frac{L}{L_o}\right) \quad (2)$$

where $M_o$ is the absolute magnitude of a reference star (e.g. the sun), $L$ is the luminosity of the radiation source at an arbitrary distance $z$, and $L_o$ is the absolute luminosity of that source[i].

Equating eqs. (1) and (2) yields:

$$z = 10^{\frac{m+5-M_o}{5}}\sqrt{\frac{L}{L_o}} \quad (3)$$

Thus, for the sun, $M_o = 4.83$, $L = L_o = 1$, and for it to be invisible to the naked eye in the [nearly] total blackness of interstellar space, $m = 6.5$ [2] (discussed in Section 4). Therefore, the sun is visible to the unaided eye at distances up to 21.58 pc (70.39 LY).

---

[i] The luminosity of the source at the absolute magnitude distance (i.e. 10 parsecs).

## 3. The Apparent Magnitude of Relativistic Blackbody Radiators

Sufficiently high speed relativistic motion of blackbody radiators would clearly Doppler shift the wavelengths of maximum luminosity to beyond the human visual range. Therefore, the lower luminosity wavelengths are Doppler shifted into the visible range, and the overall visible luminosity is reduced.

However, in the case of an approaching blackbody, the radiation is relativistically beamed, and the blackbody temperature is "inflated". Both of these effects serve to increase the luminosity. For a receding blackbody, relativistic beaming ("expanding") and temperature "deflation" will have the reverse effect. Therefore, eq. (3) becomes:

$$z = 10^{\frac{m+5-M_o}{5}} \sqrt{\frac{L'}{L_o}} \quad (4)$$

where $L'$ and $L_o$ are the luminosities in the relativistic and rest frames respectively[ii].

Since luminosity is the frequency and solid angle integrals of spectral radiance:

$$\frac{L'}{L_o} = \frac{\int_{v_1}^{v_2}\int B' d\Omega' dv'}{\int_{v_1}^{v_2}\int B_o d\Omega dv} \quad (5)$$

where $v_1$ and $v_2$ are the mean lower and upper frequencies of ocular visibility. $B'$ and $B_o$ are the spectral radiances in the relativistic and rest frames respectively, which must be integrated over the appropriate solid angle $\Omega$. The relativistic spectral radiance in frequency space, accounting for Doppler shifting, relativistic beaming, and temperature inflation, was determined by Lee and Cleaver [1], and is given by[iii]:

$$B'_v dv d\Omega = \frac{\left(\frac{2hv^3}{c^2}\right)}{\exp\left[\frac{hv}{k_B}(\beta_t - \beta_z \cos\theta)\right] - 1} [\gamma(1 - V\cos\theta)]^{-3} dv d\Omega \quad (6)$$

where [3],

$$\beta_t = \frac{1}{T_o\sqrt{1-V^2}} \quad (7)$$

$$\beta_z = \frac{V}{T_o\sqrt{1-V^2}} \quad (8)$$

---

[ii] Primed quantities indicate the relativistic frame.
[iii] For a more detailed discussion of the effects of temperature inflation and relativistic beaming on spectral radiance, see [1].

and $T_o$ is the proper absolute temperature, $u_\mu$ is the relative 4-velocity between the radiation and the observer, $\beta_\mu = \beta_t - \beta_z \cos\theta$ is the van Kampen-Israel inverse temperature 4-vector, $\theta$ is the angle between $u_\mu$ and $\beta_\mu$, $V = \frac{u}{c}$ (fraction of light speed).

The integration of the spectral radiance over all frequencies is straightforward because, with the limits of 0 and ∞, the result is simply $\pi^4/15$. However, the in-band luminosity requires integration over a finite frequency range. Here, the method of Widger and Woodall is followed [4].

$$B'_\nu d\Omega = \int_{\nu_1}^{\nu_2} \frac{\left(\frac{2h\nu^3}{c^2}\right)}{\exp\left[\frac{h\nu}{k_B}(\beta_t - \beta_z \cos\theta)\right] - 1} [\gamma(1 - V\cos\theta)]^{-3} d\nu d\Omega \qquad (9)$$

Letting:

$$Q' = \frac{\left(\frac{2h}{c^2}\right)(1 - V^2)^{\frac{3}{2}}}{(1 - V\cos\theta)^3} \qquad (10)$$

and

$$R' = \frac{h(1 - V\cos\theta)}{k_B T_o \sqrt{1 - V^2}} \qquad (11)$$

Also, letting $x' = R'\nu$, and from eqs. (7) and (8), eq. (9) becomes:

$$B'_\nu d\Omega = \frac{Q'}{R'^4} \int_{\nu'_1}^{\nu'_2} \frac{x'^3}{e^{x'} - 1} dx' d\Omega \qquad (12)$$

Expanding eq. (12) as a difference of integrals:

$$B'_\nu d\Omega = \frac{Q'}{R'^4} \left[ \int_{\frac{x_1}{R}}^{\infty} \frac{x'^3}{e^{x'} - 1} - \int_{\frac{x_2}{R}}^{\infty} \frac{x'^3}{e^{x'} - 1} \right] dx' d\Omega \qquad (13)$$

Evaluating eq. (13), and re-substituting $x' = R'\nu$:

$$B'_\nu d\Omega = \frac{Q'}{R'^4}\left[\sum_{n=1}^{\infty}\left(\frac{R'^3\nu_1^3}{n}+\frac{3R'^2\nu_1^2}{n^2}+\frac{6R'\nu_1}{n^3}+\frac{6}{n^4}\right)e^{-nR'\nu_1} - \sum_{n=1}^{\infty}\left(\frac{R'^3\nu_2^3}{n}+\frac{3R'^2\nu_2^2}{n^2}+\frac{6R'\nu_2}{n^3}+\frac{6}{n^4}\right)e^{-nR'\nu_2}\right]d\Omega \quad (14)$$

Expanding the solid angle integration, combining sums, and making use of $L'_\nu = \int_{\nu_1}^{\nu_2}\int\int B' d\Omega' d\nu'$ from eq. (5), the relativistic luminosity in frequency space (eq. (14)) becomes[iv]:

$$L'_\nu = \frac{Q'}{R'^4}\int\int\sum_{n=1}^{\infty}\left[\left(\frac{R'^3\nu_1^3}{n}+\frac{3R'^2\nu_1^2}{n^2}+\frac{6R'\nu_1}{n^3}+\frac{6}{n^4}\right)e^{-nR'\nu_1} - \left(\frac{R'^3\nu_2^3}{n}+\frac{3R'^2\nu_2^2}{n^2}+\frac{6R'\nu_2}{n^3}+\frac{6}{n^4}\right)e^{-nR'\nu_2}\right]\cos\theta\sin\theta d\theta d\phi \quad (15)$$

In the case of approaching the LAB [approximately] directly, a simplification of eq. (15), which cannot be resolved as a closed form function, can be made. Since, $\theta$ is very small $\sin\theta \sim \theta$ and $\cos\theta \sim 1$. This removes the angular dependence from eq. (11), which reduces to:

$$R = \frac{h}{k_B T_o}\sqrt{\frac{1-V}{1+V}} \quad (16)$$

Frequently, when evaluating the $d\Omega$ integration, the solid angle over which the integration is performed is the solid angle through which the blackbody radiates. However, that is not the case here. The solid angle is that which is subtended by the blackbody from the vantage point of the observer. Therefore, when $z \gg D$, $\theta \sim \frac{D}{z}$ ($z$ is the observer proper distance, and $D$ is the diameter of the blackbody). For a blackbody with a circular $x$-$y$ cross-section, $\phi \sim \frac{D}{z}$. Therefore,

$$L'_\nu \approx \frac{Q'}{R'^4}\sum_{n=1}^{\infty}\left[\left(\frac{R'^3\nu_1^3}{n}+\frac{3R'^2\nu_1^2}{n^2}+\frac{6R'\nu_1}{n^3}+\frac{6}{n^4}\right)e^{-nR'\nu_1} - \left(\frac{R'^3\nu_2^3}{n}+\frac{3R'^2\nu_2^2}{n^2}+\frac{6R'\nu_2}{n^3}+\frac{6}{n^4}\right)e^{-nR'\nu_2}\right]\int_0^{\frac{D}{z}}\theta d\theta\int_0^{\frac{D}{z}}d\phi \quad (17)$$

Thus,

$$L'_\nu \approx \frac{2k_B^4 T_o^4}{c^2 h^3}\sum_{n=1}^{\infty}\left[\left(\frac{R'^3\nu_1^3}{n}+\frac{3R'^2\nu_1^2}{n^2}+\frac{6R'\nu_1}{n^3}+\frac{6}{n^4}\right)e^{-nR'\nu_1} - \left(\frac{R'^3\nu_2^3}{n}+\frac{3R'^2\nu_2^2}{n^2}+\frac{6R'\nu_2}{n^3}+\frac{6}{n^4}\right)e^{-nR'\nu_2}\right]\left(\frac{D}{z}\right)^3\left(\frac{1-V}{1+V}\right)^{\frac{7}{2}} \quad (18)$$

Similarly,

---

[iv] The $\cos\theta$ term accounts for the Lambertian radiator, and the $\sin\theta$ term arises from the solid angle integration.

$$L_o \approx \frac{2k_B^4 T_o^4}{c^2 h^3} \sum_{n=1}^{\infty}\left[\left(\frac{R^3 v_1^3}{n}+\frac{3R^2 v_1^2}{n^2}+\frac{6R v_1}{n^3}+\frac{6}{n^4}\right)e^{-nRv_1}-\left(\frac{R^3 v_2^3}{n}+\frac{3R^2 v_2^2}{n^2}+\frac{6R v_2}{n^3}+\frac{6}{n^4}\right)e^{-nRv_2}\right]\left(\frac{D}{z}\right)^3 \quad (19)$$

Combining eq. (3), in terms of relativistic luminosity, with eqs. (18) and (19) yields:

$$z = 10^{\frac{m+5-M_o}{5}}\left[\frac{\sum_{n=1}^{\infty}\left[\left(\frac{R'^3 v_1^3}{n}+\frac{3R'^2 v_1^2}{n^2}+\frac{6R' v_1}{n^3}+\frac{6}{n^4}\right)e^{-nR'v_1}-\left(\frac{R'^3 v_2^3}{n}+\frac{3R'^2 v_2^2}{n^2}+\frac{6R' v_2}{n^3}+\frac{6}{n^4}\right)e^{-nR'v_2}\right]}{\sum_{n=1}^{\infty}\left[\left(\frac{R^3 v_1^3}{n}+\frac{3R^2 v_1^2}{n^2}+\frac{6R v_1}{n^3}+\frac{6}{n^4}\right)e^{-nRv_1}-\left(\frac{R^3 v_2^3}{n}+\frac{3R^2 v_2^2}{n^2}+\frac{6R v_2}{n^3}+\frac{6}{n^4}\right)e^{-nRv_2}\right]}\right]^{\frac{1}{2}}\left(\frac{1-V}{1+V}\right)^{\frac{7}{4}} \quad (20)$$

Evaluation of the infinite sums is greatly simplified due, in large part, to the rapid convergence of the series as a result of the $e^{-nRv}$ terms. The smallest value of $R$ (requiring the largest number of summation terms) occurs when $V = 0$, and from eq. (11), is $\frac{h}{k_B T_o}$. The smallest useful value of $Rv = 0.400$ would result from an O-class star, with a surface temperature of ~50,000 K, and at the lowest frequency of human visibility.

Table 1 gives the number of summation terms ($n$) (in eq. (20)) that would be required to produce at least 10 significant figure convergence for $0.1 \leq Rv \leq 25$.

| $Rv$ | Number of Summation Terms ($n$) |
|---|---|
| 0.1 | 101 |
| 0.2 | 65 |
| 0.3 | 50 |
| 0.4 | 50 |
| 0.5 | 35 |
| 0.6 | 30 |
| 0.7 | 25 |
| 0.8 | 22 |
| 0.9–1.4 | 20 |
| 1.5–1.9 | 15 |
| 2.0–2.9 | 10 |
| 3.0–3.9 | 8 |
| 4.0–4.9 | 6 |
| 5.0–9.9 | 4 |
| 10.0–24.9 | 3 |
| $\geq 25.0$ | 1 |

Table 1: Number of summation terms required for series convergence of eq. (20) to at least 10 significant figures [4].

## 4. The Ocular Invisibility of Relativistic Radiators

The visibility to the naked eye of astronomical objects has been discussed extensively in the literature [5], [6], [7], [8], [9]. The "sky" of interstellar space is considered to be absolutely black, and a viewing port is taken to be, at optical wavelengths, a perfectly transparent aperture that subtends a solid angle of at least the human field of vision and with a magnification of 1.

The efficiency by which photons are used by the retina was accounted for by correcting for the Stiles-Crawford effect of the first (SCE I) and second (SCE II) kind[v], photon absorption by the optical media, photopigment absorption of photons, and the photon isomerization efficiency of the photopigment.

For a 22′ (diameter), 10 ms, 507 nm monochromatic source, in which, of the ~100 quanta incident upon the retina, 10 to 15 were absorbed by the ~1600 illuminated rods [10]. From this experiment, Packer and Williams determined the rod actinometric, radiometric, and photometric[vi] absolute thresholds to be 0.35 γ/s, $4.35 \times 10^{-6}$ W·m$^{-2}$·sr$^{-1}$, and $1.33 \times 10^{-3}$ cd·m$^{-2}$ respectively [11]. However, also examined was the case of a stimulus which exceeded the visual system's spatial summation area and temporal integration time, in which the rod actinometric, radiometric, and photometric absolute thresholds are $2.00 \times 10^{-4}$ γ/s, $2.47 \times 10^{-9}$ W·m$^{-2}$·sr$^{-1}$, and $7.5 \times 10^{-7}$ cd·m$^{-2}$ respectively.

However, even accounting for the standard observer's spectral sensitivity by applying the Judd & Voss CIE 1978 photopic luminous efficiency function, these results are difficult to apply to the scenario presented here because of the enormous disparity between the spectral irradiances of the Hallett test sources [10] and stars.

The frequency range of human vision is slightly variable. However, $4.17 \times 10^{14}$ Hz and $7.89 \times 10^{14}$ Hz, which correspond to wavelengths of 720 nm and 380 nm respectively, are acceptable approximations of the limits of human vision, and are in keeping with the wavelengths of 700 nm and 390 nm published by Starr [12]. The limiting magnitude of the unassisted human eye is taken to be 6.5 [2]. This figure applies to all visible wavelengths and accounts for eye sensitivity. Consequently, inclusion of the Judd & Voss CIE 1978 photopic luminous efficiency function would not be appropriate.

### 4.1 $\theta = 0$ Ocular Invisibility

In the case of approaching the sun directly ($\theta = 0$), the distance at which the apparent magnitude is 6.5 can be determined from eq. (20), and is shown in Figure 1.

---

[v] The Stiles-Crawford effect of the first kind is the phenomenon by which light entering the edge of the pupil elicits a smaller response from the cone photoreceptors than light entering the center of the pupil. The Stiles-Crawford effect of the second kind states that the perceived hue from a monochromatic light source is dependent on its obliquity with respect to the retina.
[vi] Assuming a 6 mm in diameter pupil.

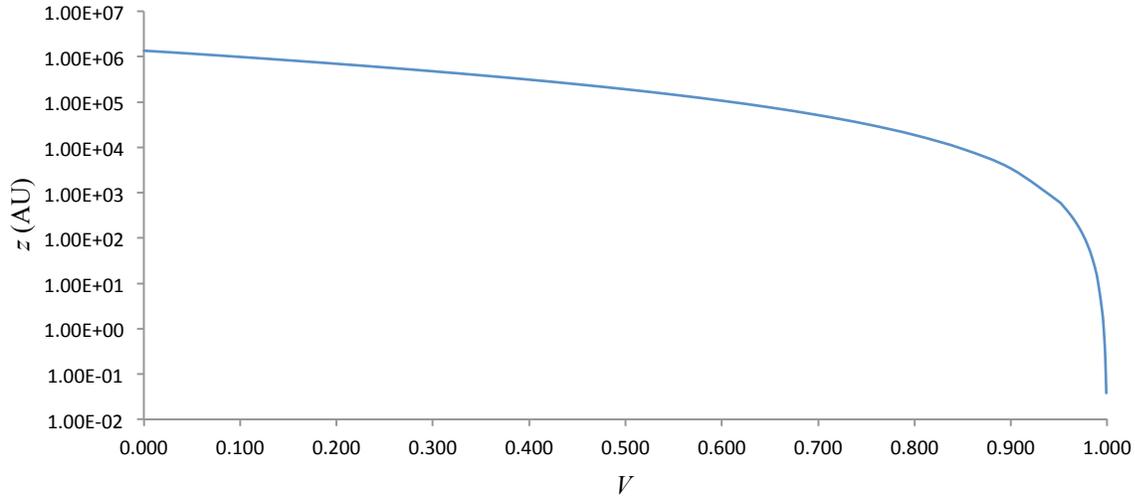

Figure 1: Distance versus speed for limiting magnitude ($m = 6.5$) of the sun. The wavelengths of vision are taken to be between 380 nm and 720 nm, and the temperature is 5780 K. The region below the curve represents the distance at which the sun is visible to the typical unaided eye of an observer in the frame of the sun.

### 4.2 Ocular Invisibility for Arbitrary $\theta$

In order to determine the ocular invisibility curve for an arbitrary velocity vector, the solid angle integration in (15), and correspondingly for the stationary case, must be performed. However, since the solid angle over which the integration must be taken does not significantly exceed ~10 mradians[vii] (and is considered primarily for angles much smaller), $z$ can be approximated as being constant at each value of $\theta$ in the 315 time step iterative scheme, which was used to evaluate the solid angle integral. When eq. (20) is evaluated for the sun, Figure 2 results.

---

[vii] The angle subtended by the sun at approximately 1 AU.

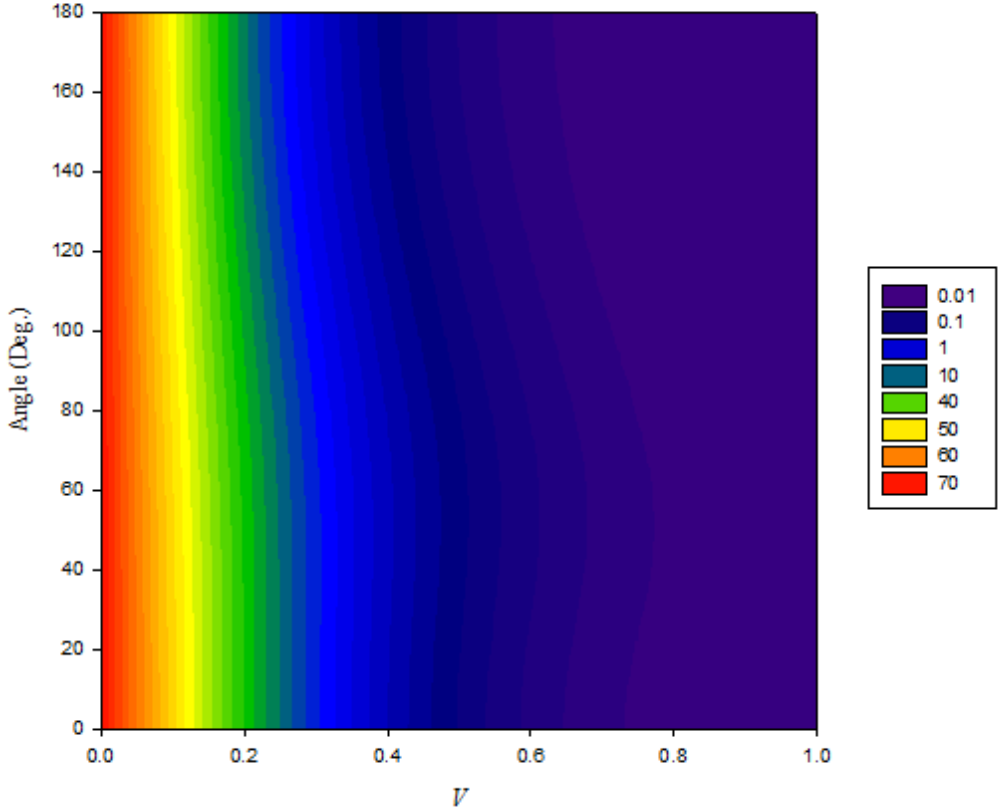

Figure 2: Proper distance of limiting magnitude as a function of $V$ and $\theta$ for the sun. The wavelengths of vision are taken to be 380 nm to 720 nm, and the temperature is 5780 K. The numbers in the legend represent the proper distances at which the apparent magnitude is 6.5. The purple region of 0.01 LY represents proper distances which are $\leq 0.01$ LY. The intersection of the contours with the *Angle*-axis is expectedly 70.39 LY (as determined in Section 2).

As expected, and shown in Figure 2, ultra-relativistic velocities permit exceptionally close approaches to luminous astrophysical bodies, while maintaining an apparent magnitude which is less than the limiting magnitude of the unaided human eye.

## 5. Conclusions

By making use of the relativistic blackbody spectrum, the velocity profile for the apparent magnitude of a LAB has been determined. Optical invisibility to the unaided eye arises due to the Doppler shifting of the wavelengths of maximum radiance to beyond the limits of human visual sensitivity. Temperature inflation and relativistic beaming can either increase this incident radiance (for an approaching source) or decrease it (for a receding source). By considering the wavelength limits of human vision to be 380 nm and 720 nm, and the limiting magnitude of the unaided human eye to be 6.5, the proper distance versus velocity function for ocular invisibility of relativistic luminous astrophysical bodies has been determined; this profile was determined for the sun.